%
% the following is to use blackboard bold fonts --
\let\useblackboard=\iftrue
%
% activate this if you don't have them.
%\let\useblackboard=\iffalse
%
% You might also need to remove this line.
\newfam\black
\input harvmac.tex
\def\Title#1#2{\rightline{#1}
\ifx\answ\bigans\nopagenumbers\pageno0\vskip1in%
\baselineskip 15pt plus 1pt minus 1pt
\else%\special{papersize=11in,8.5in}%
\def\listrefs{\footatend\vskip 1in\immediate\closeout\rfile\writestoppt
\baselineskip=14pt\centerline{{\bf References}}\bigskip{\frenchspacing%
\parindent=20pt\escapechar=` \input
refs.tmp\vfill\eject}\nonfrenchspacing}
\pageno1\vskip.8in\fi \centerline{\titlefont #2}\vskip .5in}

\ifx\answ\bigans\def\tcbreak#1{}\else\def\tcbreak#1{\cr&{#1}}\fi
\useblackboard
\message{If you do not have msbm (blackboard bold) fonts,}
\message{change the option at the top of the tex file.}
\font\blackboard=msbm10 scaled \magstep1
\font\blackboards=msbm7
\font\blackboardss=msbm5
%\newfam\black
\textfont\black=\blackboard
\scriptfont\black=\blackboards
\scriptscriptfont\black=\blackboardss

\else

\fi
% *************************************
%\draft
%
\def\yboxit#1#2{\vbox{\hrule height #1 \hbox{\vrule width #1
\vbox{#2}\vrule width #1 }\hrule height #1 }}
\def\fillbox#1{\hbox to #1{\vbox to #1{\vfil}\hfil}}
\def\ybox{{\lower 1.3pt \yboxit{0.4pt}{\fillbox{8pt}}\hskip-0.2pt}}

\def\comments#1{}

\def\half{{1\over 2}}

\def\tr{{\rm tr\ }}

\def\Im{{\rm Im\hskip0.1em}}

\def\vev#1{\langle{#1}\rangle}

\def\CA{{\cal A}}

\def\CN{{\cal N}}

\def\nl{\hfill\break}

\def\I{I}
\def\Iprime{I$^\prime$}
\def\II{\relax{I\kern-.07em I}}

\def\IIb{{\II}b}
\def\SLtwoZ{{$SL(2,\BZ)$}}

\def\IZ{\relax\ifmmode\mathchoice
{\hbox{\cmss Z\kern-.4em Z}}{\hbox{\cmss Z\kern-.4em Z}}
{\lower.9pt\hbox{\cmsss Z\kern-.4em Z}}
{\lower1.2pt\hbox{\cmsss Z\kern-.4em Z}}\else{\cmss Z\kern-.4em
Z}\fi}
\def\IB{\relax{\rm I\kern-.18em B}}
\def\IC{{\relax\hbox{$\inbar\kern-.3em{\rm C}$}}}
\def\ID{\relax{\rm I\kern-.18em D}}
\def\IE{\relax{\rm I\kern-.18em E}}
\def\IF{\relax{\rm I\kern-.18em F}}
\def\IG{\relax\hbox{$\inbar\kern-.3em{\rm G}$}}
\def\IGa{\relax\hbox{${\rm I}\kern-.18em\Gamma$}}
\def\IH{\relax{\rm I\kern-.18em H}}
\def\II{\relax{\rm I\kern-.18em I}}
\def\IK{\relax{\rm I\kern-.18em K}}
\def\IP{\relax{\rm I\kern-.18em P}}
%\def\IX{\relax{\rm X\kern-.01em X}}
%this doesn't work

\font\cmss=cmss10 \font\cmsss=cmss10 at 7pt
\def\IR{\relax{\rm I\kern-.18em R}}

\def\Im{{\rm Im\ }}
\def\BR{\IR}
\def\BZ{\IZ}
\def\BR{\IR}

\def\BP{\IP}

\def\Bone{{\bf 1}}
\Title{ \vbox{\baselineskip12pt\hbox{hep-th/9605199}
\hbox{RU-96-41}}}
{\vbox{
\centerline{Probing $F$-theory With Branes}}}
\centerline{Tom Banks}
\smallskip
\centerline{Michael R. Douglas}
\smallskip
\centerline{Nathan Seiberg}
\smallskip
\smallskip
\centerline{Department of Physics and Astronomy}
\centerline{Rutgers University }
\centerline{Piscataway, NJ 08855-0849}
\centerline{\tt banks, mrd, seiberg@physics.rutgers.edu}
\bigskip
%\centerline{others}
%\smallskip
%\centerline{xxx}
%\centerline{\tt xxx}
\bigskip
Last week, A. Sen found an explicit type \I\ string compactification
dual to the eight-dimensional F-theory construction with $SO(8)^4$
nonabelian gauge symmetry.  He found that the perturbations around the
enhanced symmetry point were described by the mathematics of the
solution of $\CN=2$, $d=4$ $SU(2)$ gauge theory with four flavors, and
argued more generally that global symmetry enhancement in $\CN=2$, $d=4$
gauge theories corresponded to gauge symmetry enhancement in $F$-theory.

We show that these $\CN=2$, $d=4$ gauge theories have a physical
interpretation in the theory.  They are the world-volume theories of
$3$-branes parallel to the $7$-branes.  They can be used to probe the
structure of the exact quantum $F$-theory solutions.  On the Higgs
branch of the moduli space, the objects are equivalent to finite size
instantons in the $7$-brane gauge theory.

\Date{May 1996}
%\draft
%
\nref\vafaf{C.~Vafa, ``Evidence for F-theory,'' hep-th/9602022.}
\nref\mvone{D. Morrison and C. Vafa, hep-th/9602114.}
\nref\mvtwo{D. Morrison and C. Vafa, hep-th/9603161.}
\nref\seiwitf{N. Seiberg and E. Witten, hep-th/9603003.}
\nref\witf{E. Witten, hep-th/9603150; hep-th/9604030.}
\nref\free{S. Ferrara, R. Minasian and A. Sagnotti, hep-th/9604097.}
\nref\aspgross{P. Aspinwall and M. Gross, hep-th/9605131.}
\nref\sen{A.~Sen, ``F-theory and Orientifolds,'' hep-th/9605150.}
\lref\gimon{E. G. Gimon and J. Polchinski,
``Consistency Conditions for Orientifolds and D-manifolds,'' hep-th/9601038.}
\lref\clp{S. Chaudhuri, C. Johnson, and J. Polchinski,
``Notes on D-Branes,'' hep-th/9602052.}
\lref\greens{B. Greene, A. Shapere, C. Vafa and S.-T. Yau,
Nucl. Phys. B337 (1990) 1;\nl
G. Gibbons, M.B. Green and M.J. Perry, hep-th/9511080.}
\lref\sw{N.~Seiberg and E.~Witten, Nucl. Phys. B426 (1994) 19,
hep-th/9407087.}
\lref\swtwo{N.~Seiberg and E.~Witten, Nucl. Phys. B431 (1994) 484,
hep-th/9408099.}
\lref\witsmi{E. Witten, ``Small Instantons in String Theory,'' hep-th/9511030.}
\lref\dbwb{M.~R.~Douglas, ``Branes within Branes,'' hep-th/9512077.}
\lref\dm{M.~R.~Douglas and G. Moore, ``D-Branes, Quivers, and ALE Instantons,''
hep-th/9603167.}
\lref\dl{M. Douglas and M. Li, ``D-Brane Realization of $\CN=2$ Super
Yang-Mills Theory in Four Dimensions,'' hep-th/9604041.}
\lref\dgdb{M.~R.~Douglas, ``Gauge Fields and D-branes,'' hep-th/9604198.}
\lref\polwit{J.~Polchinski and E.~Witten, ``Evidence for Heterotic-Type I
String Duality,'' hep-th/9510169.}
\lref\toap{M.~R.~Douglas, D.~Kabat, P.~Pouliot and S.~H.~Shenker,
to appear.}
\lref\shenker{S.~H.~Shenker, ``Another Length Scale in String Theory?'',
hep-th/9509132.}
%
% forward equation references
%
\newsec{Introduction}

$F$-theory is a powerful approach to superstring compactification,
which has been the focus of much recent work
\refs{\vafaf-\aspgross}.
Although there are strong hints that a twelve-dimensional definition of
the theory exists, at present its clearest definition is as the class of
type \IIb\ string compactifications containing $7$-branes, which
generically have non-constant dilaton and axion fields.  These fields
can be represented as the modulus of an elliptic curve fibered over the
space-time, and thus the basic structure of the compactification is
given by a complex manifold admitting an elliptic fibration.
Singularities of the fiber will correspond to positions of $7$-branes,
and encircling one of these induces an
\SLtwoZ\ monodromy on the \IIb\ fields.

The prototypical example is the representation of the surface $K3$ as an
elliptic curve fibered over $\BP^1$.  The moduli space of such
fibrations is $18$ complex dimensional and isomorphic to the Narain
moduli space of heterotic string compactifications of $T^2$.  This and
other evidence suggests that the theories are dual.  By the adiabatic
argument, it follows that compactifications on $n+1$-dimensional complex
manifolds are dual to heterotic string compactifications on
$n$-dimensional manifolds with a specific choice of gauge background.
This unification of gauge and metric parameters is a significant
advantage of the approach.

In a beautiful recent work, Sen has found an explicit heterotic dual to
an $F$-theory compactification \sen.  The starting point is the
observation that there is a special elliptically fibered $K3$ for which
the \SLtwoZ\ monodromies are almost trivial, and the dilaton and axion
fields constant -- the orbifold $T^4/\BZ_2$.  Regarding it as a $T^2$
fibration over $T^2/\BZ_2$, the $T^2$ complex structure is obviously
constant.  Thus there should be no barrier to regarding this as a
conventional type \IIb\ compactification.

In fact it is an orientifold compactification.  The \SLtwoZ\ monodromy
around a $\BZ_2$ fixed point of the orbifold as a fibration is $-\Bone$.
This has the same action on the massless fields as $(-1)^{F_L}\Omega$,
and thus we can regard the compactification as an orientifold by the
$\BZ_2$ transformation $(-1)^{F_L}\Omega R$, where $R$ is $z\rightarrow
-z$ on the base $T^2$.

This is known as a type \Iprime\ compactification \clp, in other words one
produced from the type \I\ string by $T$-duality of both coordinates of
the base.  The transformation $(-1)^{F_L}\Omega R$ is mapped to
$\Omega$, so the compactification is mapped to a conventional type \I\
theory with enhanced gauge symmetry $SO(8)^4$, $S$-dual to the heterotic
string.  It has four $7$-branes inserted at each of four orientifold
fixed points, and nonabelian gauge symmetry $SO(8)^4$.

Given the duality between $F$-theory at the orbifold point and the
calculable type \I\ theory or type \Iprime\ orientifold, it is now possible
to vary parameters in the type \Iprime\ description and learn about
$F$-theory moduli space.  Sen's next observation is that quantum effects
are always important in this problem, and that the classical
dilaton-axion solution of the form described in \greens\ is modified.

The mathematics of $F$-theory strongly suggests, and Sen argues
convincingly, that the modification is the same as the passage from the
one-loop result to the finite coupling solution of an $\CN=2$, $d=4$
$SU(2)$ gauge theory with four fundamental hypermultiplets.  The
parameters and moduli of this gauge theory are the masses $m_i$ of the
hypermultiplets and the vev $z=\vev{\tr\phi^2}$ of the adjoint scalar,
which are identified with positions of $7$-branes and a point in the
base (locally $\BR^2/\BZ_2$).  The gauge coupling $\tau$ is identified
with the dilaton-axion, and the gauge theory solution provides its
dependence on $z$ and the parameters.

In this note we point out that all this has a simple physical explanation.
One way to study the properties of a background in string theory is to study
the effective Lagrangian of an object (or `probe') moving in that background.
The background fields will turn into couplings in this Lagrangian, but
in general these are not directly observable -- only physical observables
including possible quantum corrections are meaningful.

If the background is defined by a configuration of D-branes, and the
probe is a D-brane, instead of solving for the bulk fields produced by
the background configuration, there is another way to study these
effects.  The combined system of D-branes will have various fields
corresponding to open strings stretched between pairs of D-branes.
Strings ending on the probe will correspond to fields on its
world-volume, and closed string backgrounds will produce couplings in
its Lagrangian.  Solving for the dynamics of the probe will then
reproduce the effect of the background field configuration.  This has
been demonstrated in the example of the D$5$-brane as instanton \witsmi,
by using a $1$-brane as probe \dgdb.  Further examples will appear in
\toap, along with a study of the regime of validity of this approach.

In the $F$-theory context, the $3$-brane is the natural probe, as
$7$-branes parallel to the $3$-brane will break only half its
supersymmetry, leaving an $\CN=2$, $d=4$ world-volume Lagrangian.
Furthermore, the $3$-brane is the only \IIb\ brane which is a singlet
under \SLtwoZ\ (which becomes the \SLtwoZ\ duality symmetry of its
world-volume theory), making it canonically defined in any $F$-theory
situation.  Finally, Sen's results tell us that an $\CN=2$, $d=4$ gauge
theory will naturally encode much of the physics.  The spacetime
variation of the dilaton-axion fields in F-theory is mapped onto the
variation of the low energy gauge coupling function on moduli space.  We
will see that this has the following simple physical interpretation: As
a $3$-brane is moved around in spacetime, the coupling of the gauge
theory on its world volume is determined by the background fields.  The
adjoint VEV which parametrizes the moduli space of the world volume
gauge theory {\it is} simply the $3$-brane position, relative to the
orientifold point of the type \Iprime\ theory.
Furthermore, it is now easy to understand the
appearance of $SU(2)$ gauge theory.  The $5$-brane of type \I\ theory
comes equipped with $SU(2)$ Chan-Paton factors \refs{\witsmi, \gimon}.
By T-duality these are inherited by the type \Iprime\ $3$-brane.
The five brane T dual to our $3$-brane probe is wrapped around the two
torus. A Wilson line in its world volume $SU(2)$ gauge theory breaks the
gauge group down to $U(1)$.  In the T dual theory, the $3$-brane lives
in the noncompact dimensions and can move around on the torus.  $SU(2)$
symmetry is restored only when it approaches one of the orientifold
points.  For other configurations, the massive charged gauge bosons are
realized as open strings connecting the $3$ brane to the orientifold point.

A final aspect of Sen's result which is clarified by our analysis is the
fact that the gauge symmetries of string theory showed up as global
symmetries of the $SU(2)$ gauge theory which describes the curve. Gauge
symmetries always appear as global symmetries in the effective
lagrangian of an extended object.  This is necessary for consistent
coupling of that object to background gauge fields.

\newsec{Type \Iprime\ orientifold and $F$-theory near an orbifold point}

We start with the equivalent type \I\ theory on $T^2$ with gauge
symmetry broken by Wilson lines with holonomy $\exp 2\pi
i\sum_i\alpha_{i,\mu}R_i$, with $1\le i\le 16$, $\mu\in\{8,9\}$ and
$R_i$ generators of $U(1)^{16}\subset SO(32)$.  Explicit Wilson lines
breaking to $SO(8)^4$ are $\alpha_{4j+i,8}=j/2$ and
$\alpha_{8k+i,9}=k/2$.  The complete gauge symmetry includes $U(1)^4$ as
well {}from the metric and $C^{(2)}$ fields on $T^2$.

Introduce a $5$-brane transverse to the $T^2$, the $T$-dual of the
$3$-brane of the introduction.  Its world-volume theory is $\CN=1$,
$d=6$ $Sp(1)$ gauge theory, which on $T^2$ has Wilson line moduli
$\CA^a_\mu$ with eigenvalues $\theta_\mu$.  It also contains $32$
$Sp(1)$ doublet `half-hypermultiplets' \witsmi.  These will be given
mass by the Wilson lines.  The matter can be rewritten as $16$ doublet
hypermultiplets with masses
$(\alpha_{i,8}+i\alpha_{i,9})-(\theta_{8}+i\theta_{9})$.  The maximal
massless matter content is $4$ hypermultiplets attained by taking
$\theta_{8}=\theta_{9}=0$.

The $F$-theory limit is the limit of small $T^2$, in which momentum
modes in the $T^2$ decouple.  After $T$-duality of both of these
dimensions the Wilson line parameters become positions on the dual
$T^2/\BZ_2$.  Call the complex positions of the $7$-branes
$m_i=(R'_8\alpha_{i,8}+iR'_9\alpha_{i,9})$ and the complex position of
the $3$-brane $A=A^a\sigma_a=(R'_8\CA_{8}+iR'_9\CA_{9})$.  This can be
diagonalized to
\eqn\higgsm{A=\left(\matrix{w& 0\cr 0& -w}\right),}
the expectation value of an adjoint Higgs breaking $Sp(1)$ to $U(1)$.
The gauge-invariant parameter is $z=\half\tr A^2=w^2$.

As in \gimon\ it is convenient to work on the covering space $T^2$,
introducing a $\BZ_2$ image for each brane.  From now on we focus on
the neighborhood of a single $\BZ_2$ fixed point, $w=0$, and
ignore the $7$-branes far from the fixed point.
The images are then located at $-m_i$ and $-w$.
The pair of $3$-branes at $(w,-w)$ are the same pair introduced in
type \I\ theory to get $Sp(1)$ and these images lead to the structure \higgsm,
while the two $7$-brane images each give open strings in half-hypermultiplets.
At tree level, the half-hypermultiplets have the ($\CN=1$) superpotential
\eqn\hhsup{W = \sum_i m_i h_i \tilde h_i + \tr h_i A \tilde h_i}
and obtain masses $m_i+w$ and $m_i-w$.

In the gauge theory context, the next step is to compute the perturbative
renormalization in the $U(1)$ theory with charged matter.
This produces the effective gauge coupling
\eqn\tauoneloop{
\tau(z) = \tau_0 + {1\over 2\pi i}
	\left(\sum_{i=1}^4 \log(z-m_i^2) - 4\log z \right).
}
The cutoff would a priori be string scale but in fact cancels in this
finite theory.  This is clear from the alternate interpretation of the
result as the dilaton-axion produced by the $7$-brane sources, probed by
the $3$-brane.

As Sen points out, this result is physically unacceptable as it does not
satisfy $\Im\tau\ge 0$ everywhere in the complex plane.  At strong
coupling, D-instanton corrections, which behave as $e^{-\tau}$, will
become large.  These are equivalent to the $3$-brane gauge theory
instantons and the exact quantum result for $\tau$ is that given in
\swtwo.

This leaves the question of why the masses of open strings stretching
{}from one $7$-brane to another are given by the simple formulae $m_i\pm
m_j$.  Properly speaking, this is a question of $7$-brane quantum
dynamics, but we can infer the answer from our results by a simple
trick.  The strings stretching from the $3$-brane to a $7$-brane are
just the matter multiplets whose masses are given by $m_i\pm w$.  Now
take the $3$-brane to the point $w=m_j$.  The masses of these strings
will then be $m_i\pm m_j$.  But a $37$ and a $77$ string are both BPS
saturated, and both couple to the bulk fields in exactly the same way.
The central charge and therefore the mass must be the same for both.  We
have thus reproduced all of Sen's results from the physics of $3$-branes.

\newsec{Discussion}

Sen has made a link between $F$-theory and the quantum dynamics of
$7$-branes.  We used a $3$-brane to see these quantum effects, and
argued that the quantities studied by Sen, the dilaton-axion and BPS
masses, had observable counterparts on the $3$-brane.  However the
effects would be present without the $3$-brane.  This leads to the
question: can we understand the quantum dynamics of the $7$-branes more
directly?

The same object, the D-instanton, is responsible for the
non-perturbative corrections in both $3$-brane and $7$-brane theories.
In the $3$-brane context it is equivalent to the gauge instanton, whose
effects are known.  The low energy field theory of the $7$-brane is
not renormalizable, but perhaps nonrenormalization theorems would allow
us to perform an exact field theoretic computation of certain
quantities.

The $3$-brane coupling $\tau$ can go to zero at certain points, and
naively one might have interpreted this as the tension going to zero and
the $3$-brane becoming light.  However, one should do an \SLtwoZ\
transformation to weak coupling to see the real situation.

The approach is clearly very general.  One can repeat the analysis for
general D-brane configurations at general orbifold singularities to get
more general matter.  Qualitative properties of the $3$-brane gauge
dynamics will translate directly into properties of the geometry.  For
example, with more than four $7$-branes at the fixed point, the theory
will be IR free, and the orientifold point will not split.

When several 7-branes coincide the eight dimensional space time theory
has enhanced non-Abelian gauge symmetry, and thus has Yang-Mills instanton
solutions.  These will be BPS saturated $3$-branes which are expected
to be equivalent to the Dirichlet $3$-brane on an appropriate branch of
its moduli space \dbwb.
Indeed, the 3-brane
theory has a corresponding enhanced global symmetry at these points,
and a Higgs branch.   Moving along the Higgs branch, the instanton size grows.
This interpretation can be justified in several different ways.  For
example, in the type I theory, a space time instanton is on the Higgs
branch of the D5-brane.  As this brane wraps the two torus, it leads to
our 3-brane whose Higgs branch is this instanton.  $\CN=2$ supersymmetry
implies that the metric on this branch does not receive quantum
corrections
\nref\aps{P.C. Argyres, M.R. Plesser and N. Seiberg,
hep-th/9603042.}%
\refs{\swtwo, \aps}.

In this context, it will be interesting to search for gauge theories
with $E$-series global symmetry, to explain the $E_8$ gauge symmetry of
the heterotic string.  This symmetry might arise non-perturbatively at
special points in moduli space.  We have to vary the base to get these
symmetries, and a logical place to search for it is at other orbifold
points $\BR^2/\BZ_n$.

One could also introduce more than one $3$-brane, to get more general
gauge groups.  We could no longer regard a single $3$-brane in this
system as an isolated probe, but clearly this is a very simple way to
produce interesting $d=4$ gauge theories.  There are a host of further
generalizations of these ideas. For example, we can consider  type \I\ on
$K3$ and wrap a five brane on a two cycle.  The resulting 3-brane in six
dimensions has $\CN=1$ supersymmetry.  There can be an interesting
interplay between the dynamics of this theory and the background it
propagates in.

\medskip
\centerline{Acknowledgements}
This work was supported in part by DOE grant DE-FG02-96ER40559
and NSF grant PHY-9157016.  We thank A. Sen, S. Shenker and E. Witten
for discussions.

\bigskip

\listrefs
\end